# New Statistical Methods For Systematizing The Nuclei Fission Fragments: Post-Scission Approach

V. T. Maslyuk, O. A. Parlag, O.I. Lendyel, T. I. Marynets and M.I. Romanyuk

*Institute of Electron Physics, National Academy of Sciences of Ukraine, Uzhgorod, Ukraine*
*e-mail: nuclear@email.uz.ua*
*volodymyr.maslyuk@iep.org.ua*

**Abstract:** A new statistical method for systematizing the nuclei fission fragments, investigation of their mass and charge spectra and neutron fission parameters has been suggested. In proposed method, the mass and charge yields of nucleons are determined from the equilibrium conditions of the nuclear fragments post-fission ensemble. Here we consider protons and neutrons of different fragments as statistically non-equivalent. The abilities of the proposed statistical method for explanation of the U, Pa, Th, Ac and Ra post-fission fragments ordering and neutron emission function calculation have been demonstrated.

## 1. Introduction

It is known that the anisotropy of the nuclei fragments yields is common, but strongly expressed result of the atomic nuclear fission [1–4]. Information on characteristics of this phenomenon is important for understanding the nature of the internucleon forces, the roles of the shell effects and the magic nuclear numbers in the fission channels formation. At the same time, the studies of the mass and charge spectra of fission fragments (MCSFF) formation are important for a wide spectrum of applied applications ranging from nuclear geochemistry, radioecology to nuclear power engineering and medicine.

The most significant features of the MCSFF are:
1. symmetric (one-hump) or asymmetric (two- or three-humps) structure of the MCSFF;
2. MCSFF determined via the composition of the initial nucleus (the numbers of the protons and neutrons);
3. at the same excitation energy the MCSFF weakly depends on the the nucleus fission channel of (γ, f), (α, f), (f, f'), (n, f) or (p, f) [5];
4. strong dependence on the excitation energy or nucleus temperature (T), namely, it becomes more symmetric with the rise of the T.

A number of theoretical approaches treat the problem of the nuclear fission based on the particles scattering theory, the change of the nucleus shapes and the profile of its potential energy, fission dynamics, formation of the mass channels, etc. Conditionally, they can be formally divided into four groups [6–12]:
- Statistical: the thermodynamics equilibrium state is valid before the atomic nucleus fission.
- Adiabatic: the nucleus deformation is smaller as compared to the nucleons motion.
- Empirical: the fission fragments yield modes tabulation etc.
- Computational: the methods implemented in the codes like ABLA, TALYS, etc.

Despite a big progress achieved with these methods in calculating the nuclear fission transformations, the traditional methods of the MCSFF investigation requires a number of adjustable parameters such as effective viscosity, surface tension, level densities, etc.

Moreover, the investigation of the nuclear fission needs to take into account stochastic effects that are typical for small systems [13]. The *ab initio* calculations are also impossible because there is no sufficient knowledge on the nature of internuclear forces.

A series of experiments on the 1A GeV scale accelerators also confirmed the actuality of studies of the nuclear fission fragments ordering. First of all, it is the well known Darmstadt experiment (DE) [14] resulted in the complex form of the charge spectrum for 46 isotopes of U, Pa, Th, Ac and Ra photofission.

Then, the experiments on multifragmentation of heavy nuclei with the energies 1-10 MeV per nucleon showed the possibility to apply classical approaches of condensed and gaseous states of matter to nuclear subsystem and interpretation of nuclear fission in terms of phase transitions „liquid-gas" [15]. Such studies are related to a new approach named nuclear thermometry. It allows one to treat the transformation of the nuclear matter in the terms of thermodynamics or chemical methods.

In our previous works [16, 17] we suggested the statistical approach (color statistics), where the subject of the studies is the post-scission state of the nucleus derived from the thermal equilibrium condition of final nuclear fragments ensemble. This method allows to use statistical thermodynamic methods to investigate the ordering of the nuclear fission fragments, see also [18]. The first result of using this approach was interpretation of the experimental data on the Xe and Kr isotope yields from light actinide fission [19].

In the present paper, other abilities of this method are demonstrated. The structure of the paper is following: chapter 2 is devoted to the explanation of the proposed statistical method; in chapter 3 we present the results of nuclear fission characteristic calculations (MCSFF, neutron function) on the example of $^{236}$U and DE; chapter 4 is the conclusive one.

## 2. Theory

In the proposed statistical method, the subject of investigation is not the pre-, or scission state of the initial nucleus, but a system of it fission fragments with very peculiar thermodynamical properties, their ordering and hierarchy. It is well known about specificity use of thermodynamics for small systems. Following paper [20], the concept of nuclear temperature $T$ for a small nuclear system is valid in cases when the temperature fluctuation is $\Delta T/T \ll 1$, where

$$\Delta T/T = 2/\sqrt{AT} = (2MeV/AU)^{1/4}.$$

Here U is an internal energy of nucleus, A is the atomic mass. Estimates show that the parameter T can be applied for nuclear systems with $A \geq 80$.

### 2.1 Fission fragments as a thermodynamic ensemble

We consider the scheme of the two-fragment fission and than must to realize all possible distribution of nucleons of initial nucleus by two fragments with their different atomic masses and the protons/neutrons ratio. Each such distribution creates the two-fragment clusters and the set of fragment clusters form a statistical ensemble.

In this case, the $i$-th cluster contains $Z_{j,i}$ protons and $A_{j,i} - Z_{j,i}$ neutrons in the $j$-th fragment, where $j = 1,2$, as well as $n_i$ fission neutrons. In general case, the following conservation conditions for all possible schemes of two-fragment fission hold true

$$A_{1,i}+A_{2,i}+n_i=A_0, \quad Z_{1,i}+Z_{2,i}=Z_0 - \Delta z_i, \quad (1)$$

where $\Delta z_i$ is the number of β$^+$ (at $\Delta z_i$ <0) or β$^-$ ($\Delta z_i$ >0) decays within a single nuclear cluster. Emission of nuclear particles plays an important role in the relaxation of the excitation of heavy-nuclei fission fragments and their approaching the islands of stability

The thermodynamics parameters of the ensemble of two-fragment clusters are determined by the state of the initial nucleus. The initial nucleus determines the type of created statistical ensemble by considering the fluctuations of one set of the thermodynamics parameters and neglecting the other. In case of nuclear fission we assume the number of nucleons of all types in the two fragment clusters is constant, see (1) and only fluctuation of energy and volume is allowed. This leads to a canonical constant - pressure (P) ensemble. For example, in case of nuclear fusion as a result of the two nuclei interaction [18], the conception of Gibbs microcanonical ensemble can be proposed.

Since time and causality are not the parameters of the thermodynamic method the post scission evolution of fission fragments could be taken into account by the introduction of a sets of the ensembles of fragments clusters containing the nucleus restricted by their half-life $T_{1/2}$. For all this ensembles we would assume that the equilibrium conditions are valid. The experiment, indeed, shows that the fragments separate at the scission point with an almost constant temperature [21].

**2.2 The thermodynamic functions of the fission fragments ensemble**

Within the proposed approach, the problem of studying of fission fragments yields is reduced to the analysis of the equilibrium conditions for a canonical constant-pressure ensemble. We must take into account that the emission of fission neutrons decrease of the nucleus volume [22] and thus provides the $P\Delta V$ work. The equilibrium parameters of the two-fragment clusters ensemble can be obtained from the condition of minimum of the Gibbs thermodynamic potential [23]:

$$G = U - TS + PV, \quad (2)$$

here $T$ in MeV, $U$, initial or total energy consist from two major components, kinetic and potential energy. The kinetic energy is due to the motion of the system's particles and is constant in our case because of constant temperature, T. The potential energy is associated with the static constituents of matter and for condensed nuclear matter as a nucleolus, $U$ determined by binding energy of the two-fragment cluster and its spectrum $\{\varepsilon_i\}$ is an additive quantity with respect to the binding energy of fission fragments:

$$\varepsilon_i = \sum_{j=1,2} \cdot \sum_{\langle N_p \rangle_i} \cdot \sum_{\langle N_n \rangle_i} U_j(A_{j,i}, Z_{j,i}), \quad (3)$$

where $U_j$ is the binding energy of the $i$-th fission fragment, $j = 1, 2$; the symbol $\langle...\rangle_i$ means that summation in (3) is taken over the $i$-th clusters containing two fission fragments with the numbers of protons ($N_{j,i}^p$) and neutrons ($N_{j,i}^n$) satisfying the following condition:

$$\sum_{j=1,2} (N_{j,i}^p + N_{j,i}^n + n_{j,i}) = A_0. \quad (4)$$

The configurational entropy $S_i$ introduced in (2) is determined by the degeneracy factor $\omega_i$ or number of states two-fragment cluster which have the same energy level $\varepsilon_i$ and must take into account the statistical nonequivalence of nucleons with different specific binding energy in each of the fission fragments and fission neutrons, $S_i = \ln(\omega_i)$, where:

$$\omega_i = A_0! K(n_i) / (\prod_{j=1,2}(N_{j,i}^p! N_{j,i}^n!)). \quad (5)$$

Here $\prod_{j=1,2} x_j! = x_1! x_2!$ and $K(n_i)$ depend on the statistical equivalence of emitted neutrons,

| $K(n_i)=$ | | |
|---|---|---|
| | $\dfrac{1}{(n_{1,i}+n_{2,i})!}$ | emitted neutrons are statistically equivalent. |
| | $1,$ | emitted neutrons are not statistically equivalent. |
| | $\dfrac{1}{n_{1,i}! n_{2,i}!}.$ | Emitted neutrons are statistically equivalent within a single fragment, $n_{1,i}+n_{2,i}=n_i$. |

where $n_{1,i}$ ($n_{2,i}$) is the number of neutrons emitted from the first (second) fission fragments and the $i$-th cluster.

From the analysis of expression (5) we see that the entropy term in Eq (2) reaches maximum if $N_{1,i}^p = N_{2,i}^p$, $N_{1,i}^n = N_{2,i}^n$ and is responsible for the symmetrization of the fission yields with the rise of the nuclear temperature $T$.

The isobaric distribution function, which describe the statistical properties of a system in thermodynamic equilibrium and represents a probability of finding a two-fragment nuclear cluster in the $i$-th state of the ensemble with the energy $\varepsilon_i$, can be expressed in the following way:

$$f_i(V) = \omega_i \exp\{-(\varepsilon_i + PV)/T\}/Z_p, \quad (6)$$

where the statistical sum $Z_p$ is defined as:
$$Z_p = \sum_{k,V} \omega_k \exp\{-(\varepsilon_k + PV)/T\}.$$

The set of equations (2)–(6) is sufficient to study the observable characteristics of nuclear fission, including MCSFF.

## 3. The statistical properties of an post-scission fragments ensamble and observable characteristics of nuclei fission

The abilities of the proposed statistical method are demonstrated in the following calculations of the observable post-scission characteristics of nucleus fission, mainly on the example of $^{236}$U, which has the most complete nuclear-physical database [24, 25]. The statistical fluctuations of the thermodynamics parameters typical for small nuclear systems were also taken into account.

First, we investigate the MCSFF yields. This was carried out using two sets of the ensembles of fragments clusters restricted by their half-life $T_{1/2}$: the most complete database of possible fission fragments (from ultra short-lived to long-lived) and the same after exclusion the short-lived isotopes. The last case is capable for reconstructing the real experimental conditions of the MCSFF measurements.

Second, since the theory takes into account the presence of neutron fission in the two-fragment fission scheme, it is possible to calculate the number of equilibrium neutron as a function of the initial fragment mass $\nu(A)$ and the total neutron multiplicity $\bar{n}$.

And, finally, we offer own interpretation the DE results [14] that demonstrated the symmetrical and asymmetrical structure of the light actinide nuclear charge spectra of the fragment fission.

### 3.1 MCSFF post-scission characteristic and their temperatures ordering

To study MCSFF and other experimentally observable parameters one has to start from equation (5). Then, the distribution function $F(A_1)$ of a single fission fragment with mass $A_1$, or the same, $F(Z_1)$, with charge $Z_1$ has to be obtained by a following procedure:

- forming the whole ensemble of post-scission fragment clusters, using for nucleons conservation conditions (1), (3);
- the initial (not normalized) values $F(A_1)$ are obtained as the sum of probabilities of two-fragment cluster, containing the fission fragment with the mass $A_1$, see Eq. (5). This procedure is similar to the method of histograms and must includes cumulative chains;
- the same procedure is valid for determination the initial values of $F(Z_1)$;
- the Monte Carlo procedure must be applied to simulate the statistical fluctuations of the thermodynamical parameters of the fission fragments ensemble;
- the next step includes the normalization procedure and determination of the final values of $F(A_1)$ and $F(Z_1)$. These functions must satisfy the following normalization equations:
$$\sum_{<A_1>} F(A_1) = \sum_{<Z_1>} F(Z_1) = 200\%,$$
where $<A_1>$, $<Z_1>$ means the same as in (3).

In the proposed statistical method, the isobaric term $PV$ (see Eq. (6)) was chosen in the form
$$PV = P(V_0 - v_0 n),$$
where $V_0$ is the initial nucleus volume, $P$ is the nucleon "gas" pressure and $v_0$ is the average volume related to a single fission neutron with total number of $n$. The isobaric constant $Pv_0$ value was estimated within 4 – 5 MeV and was evaluated from the condition that the total fission neutron number $\bar{n}$ does not exceed 3 neutrons per fission (at T~0.5 MeV). Estimation shows that a $Pv_0 n$ value is less than 3% of the total $U$ value in Eq. (3).

It should be noted that proposed statistical method contains no adjustable parameters, but only those that can be obtained from experiment. For instance, the temperature $T$ can be defined by analyzing the evaporation spectra of the fission neutrons/protons or fission fragments (see [21, 26, 27] for more detail). The binding energies used in Eq. (3) are tabulated in [28], their extrapolation (mass formula) is given, for example, in [29–31].

Since the theory does not allow to take into account the time evolution of ensembles of fragments clusters, an appropriate choice of nuclei sets $\{A_j, Z_j\}$ can be used for better fitting the calculated and experimental MCSFF data. To reconstruct the ideal and real experimental conditions of the MCSFF measurements one can use the ensembles of fragments clusters restricted by their half-life $T_{1/2}$. Such ensembles are constructed on nuclei sets $\{A_j, Z_j\}$ that satisfy conservation conditions (1), (4) and obtained by following way:

- *the most complete sets $\{A_j, Z_j\}$* that contains all, from the ultra short-lived to long-lived and practically unobserved nucleons. The parameters of such sets as spectrum $\{\varepsilon_i\}$ can be evaluated only on the basis of the mass formula [29–31];
- *the observable nuclei sets $\{A_j, Z_j\}$* that contain observable and tabulated nucleus fragments, e.g. in Ref. [28];
- *the nuclei sets $\{A_j, Z_j\}$* containing no short- and ultra short-lived nuclei.
- *the same nuclei sets $\{A_j, Z_j\}$*, but with taking into account the particle emission, such as neutrons or beta-particles.

Figure 1 presents the experimental [25] and calculated MCSFF (yields) for the $^{236}$U. The theoretical curves were obtained for the two sets of the ensembles of fragments clusters, restricted by their half-life $T_{1/2}$: the observable nuclei sets from database [28] for 1848 nuclei and the same after exclusion the short-lived isotopes with the half-life, $T_{1/2}$<5 min.

The following parameters were used in our calculations: T=0.5 MeV, the isobaric constant $Pv_0$ was equal to 4.5 MeV and the fluctuation range was up to 15% for all $\Delta T$, $\Delta(Pv_0)$ and for the neutron emission number $\Delta(n_j)$. The number of statistical events was 100 and the length of the cumulative chain was 10. The entropy term was calculated under the assumption that all emitted neutrons are statistically equivalent, $K(n_i) = \dfrac{1}{n_i!}$ (see Eq. (5)).

The second hump for heavier fragments in the theoretical mass spectra, Fig. 1a) is related to a cluster, which contains heavy fragments in the vicinity of $A_2$=132 or the nucleus with magic $Z_2$=50 and $N_2$=82. Our theoretical data cannot describe well the experimental mass spectrum in the vicinity of $A_1 = 90$ and $A_2 = 140$ for the mass fragment yields. It can be caused by the singular tendency of the isobaric distribution function for two-fragment cluster Eq. (5) that may significantly affect the dependences $F(A_1)$ or $F(Z_1)$ for single fragments.

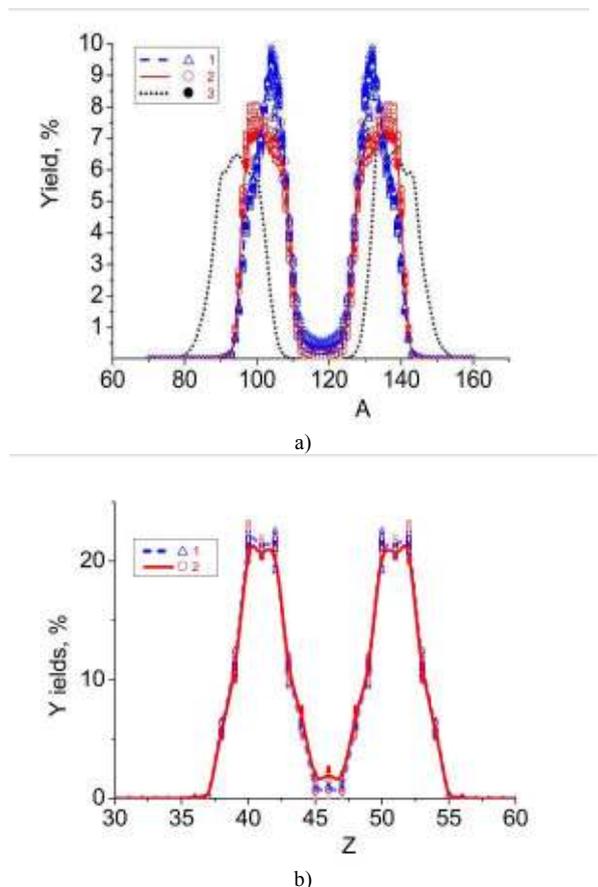

b)
Figure 1. (Color online) Mass (a) and charge (b) spectra of the $^{236}$U fission fragments obtained by using database from Ref. [28] (triangles and dashes curves) and in case of neglecting the short-lived nucleons ($T_{1/2}$< 5 min.) (circles and solid curves). The dotted curve in Figure 1(a) is related to the experimental fission fragment yields (spectrum) for the $^{236}$U: fission induced by the thermal neutrons, ($n_{th}$, f) $^{235}$U [23].

Figure 1b shows the nuclear charge spectra $F(Z_1)$ for the $^{236}$U fission fragments, normalized to 200%. The theoretical curves 1 and 2 are related to the same sets $\{A_j, Z_j\}$ as for the mass spectra in Figure 1a. Similarly to the previous case, one can see that the maximum of the charge spectra (cumulative yields) is formed by the fission fragments from the vicinity of a magic number, in particular, $Z_1 = 50$, for heavy fragment. However the experimental charge spectrum [14] indicates the location of the maximum of heavy fragments yields in the vicinity of $Z_2 = 55$. As in the previous case this could be caused by the singular tendency of the isobaric distribution function for two-fragment cluster Eq. (5) or that the emissions of an β-particles was not being taken into account.

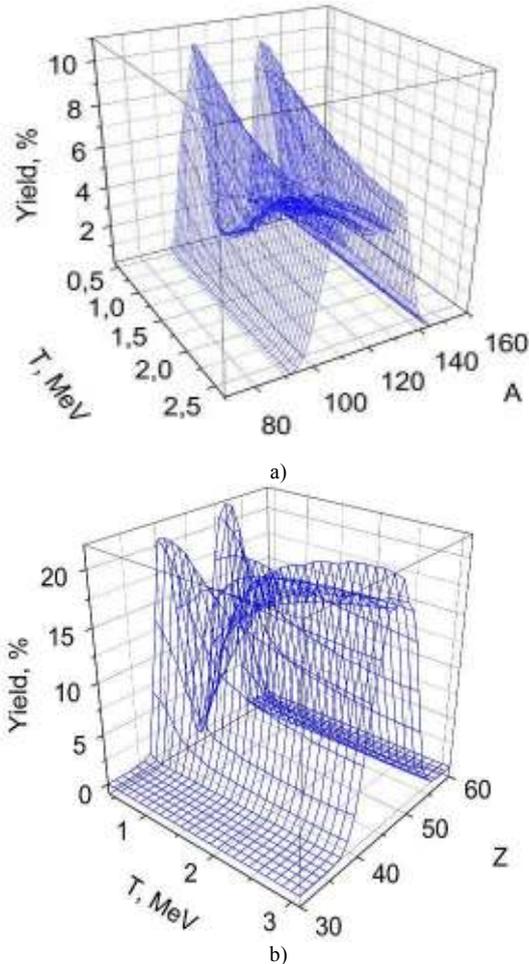

Figure 2. (Color online) Symmetrization of the nuclear mass (a) and charge (b) spectra with the temperature for the $^{236}$U fission fragments.

But in both cases in interpreting the MCSFF (yields) the better agreement between experimental and theoretical results are achieved when we used the nuclei sets {$A_j$, $Z_j$} with long-lived nucleons and taking into account the post-scission nuclear particle emission. This shows the importance of experimental conditions in the MCSFF studies.

In Figure 2, the temperature evolution of the $^{236}$U MCSFF for $Pv_0 = 4.5$ MeV on the basis of a most complete set of {$A_j$,$Z_j$} [28] is shown. The effect of $F(A_1)$, $F(Z_1)$ symmetrization with the rise of the nuclear temperature T can be explained by the influence of the entropic term in Eq. (2), which reaches a maximum in case of symmetric fission, when $A_1 \sim A_2$ ($Z_1 \sim Z_2$).

The theoretical results in Figure 2 show that the parameterization of the anisotropy of the nuclear mass (charge) spectra of the fission fragments is very promising for developing a new type of nuclear thermometers.

### 3.2. Neutron emission function for the $^{236}$U fission fragments

As was mentioned above the theory allows to calculate some observable neutron parameters: the total number of emitted neutrons as a function of the initial fragment mass (neutron emission function) $v(A)$ and total neutron emission number $\bar{n}$.

These functions are very important for neutron physics and numerous applications of neutron fluxes [32]. Among the factors that determine $v(A)$ and $\bar{n}$, the isotopic composition of the initial nucleus and its excitation energy or temperature $T$ are the most important [33]. According to Eqs. (3) and (4), the emission of neutrons from fission fragments changes both the total binding energy of a two-fragment cluster and its entropic term.

The method of the $v(A)$ function calculation is based on determining the probability of realization (yield) of the two-fragment cluster that contains a preneutron fragment with mass $A$ and the equilibrium number of neutrons $n$. Considering a cumulative yield of the fission fragments, $\bar{v}(A)$ is equal to a mean value for all clusters containing the fragment with mass A from the cumulative chain:

$$\bar{v}(A) = 1/m \sum_{j=1}^{m} v_j(A),$$

here m is the length of the cumulative chain that forms the yields of the fragments with the mass A. The total number of neutrons emitted in the act of nucleus fission $\bar{n}$ is calculated in a following way (normalization to 200% is used):

$$\bar{n} = 1/200 \sum_{A_i=1}^{A_i=A_0} v(A_i) F(A_i).$$

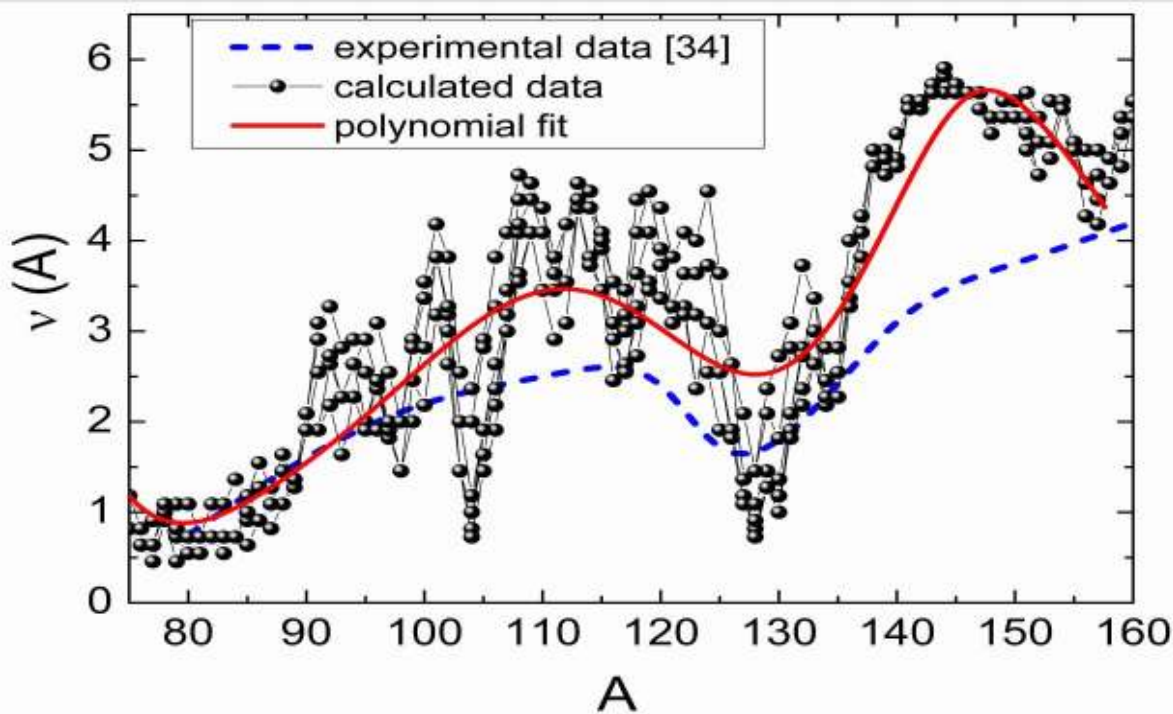

Figure 3. (Color online) The fission neutron yield is given as a function of the fusion fragment mass of $^{236}$U: dashed line corresponds to the experimental data [34], circles with line present the calculated data and the solid line is the polynomial fit of our data (see more detail in the text).

In Figure 3, the calculated and experimental [34] neutron emission functions for fission fragments of $^{236}$U are shown. As in previous cases the cumulative chain length was $m=10$. The values of T, $Pv_0$, their fluctuation varies and the number of statistical events were the same as in the previous calculations described above. We obtained that the average value $\bar{n}$ for such number of statistical events is 2.69 particles.

As one can see, the theoretical data agree well with the experiment. Moreover, the 9th order polynomial fit of our data (solid line in Figure 3) reproduces the known experimental "sawtooth"-curve of the neutron multiplicity, namely the peak about 115, minimum in vicinity of 128, the further growth in the range of 145 and decrease to 160. In addition, the proposed statistical method allows one to obtain the fine structure of $\nu(A)$, like local minima at 98, 104, 117, 121, 128, 134, 158 and local maxima at 92, 101, 109, 114, 119, 124, 133, 144, 154.

It should be noted that the experimental dependences $\nu(A)$ provide no data on such fine structure [34-38]. The fine structure might be caused by many factors as the presence of light and heavy fragments with the magic and near-magic numbers in the cluster, by the optimal proton/neutron ratio in the fragments or by the influence of the odd/even effects, etc.

### 3.3. Calculation of the nuclear charge spectrum of the U, Pa, Th and Ac isotopes

The charge spectra of the U, Pa, Th and Ac isotopes were studied in the experiment with radioactive beams formed by the $^{238}$U fragmentation at the DESY Darmstadt heavy-ion synchrotron SIS (see [14,18, 39] for experimental details). In particular, the charge spectra of the $^{230/234}$U, $^{223/232}$Pa, $^{220/229}$Th and $^{219/226}$Ac isotope series were investigated. They have different abilities of nuclear transformation because of a significant variation of their specific binding energy, ε (fissionability parameter, $Z^2/A$ ). Namely, for the U series it is 7.6209/7.6007 (36/36), for the Pa series it is 7.6519/7.5946 (37/35), for Th it is 7.6846/7.6349 (36/35) and for Ac it is 7.7006/7.6556 (36/35). These isotopes in the series also differ drastically by their half-live periods, for example, in the Th series – from $10^{-6}$ s to $7 \cdot 10^3$ years!

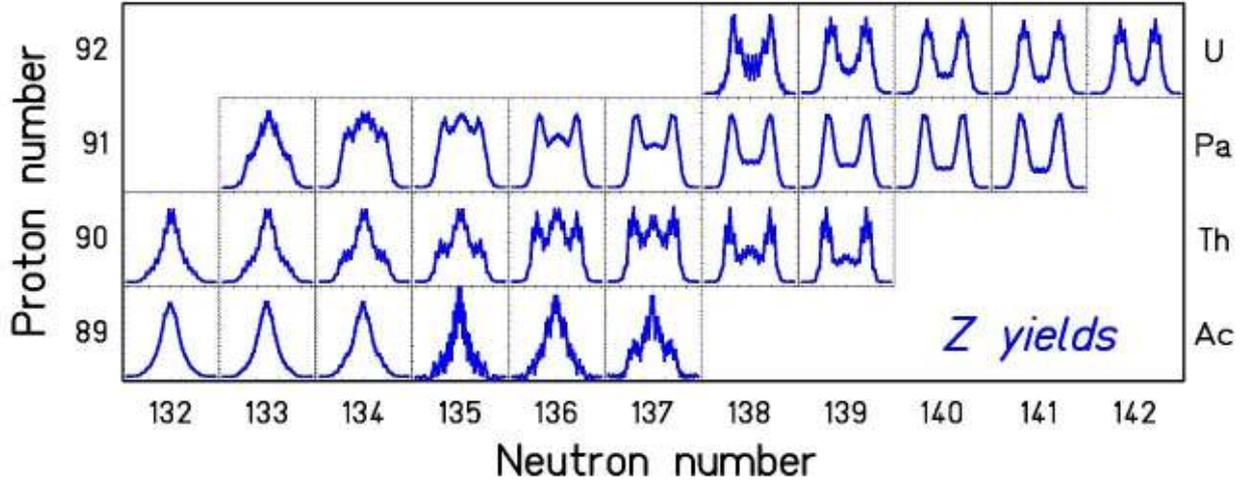

Figure 4. (Color online) Experimental fission-fragment nuclear-charge spectra of the U, Pa, Th, Ac isotopes, taken from [14]

The experimental data (see Figure 4) indicate the different shapes of the charge spectra for the isotopes of the same chemical compound. In our method, the main features of the DE experiment can be explained by by letting the possibility of varying temperature T of the initial isotopes within 0.7 - 0.9 MeV at fixed values the set of basic model parameters: isobaric constants, $Pv_0$, the emitted neutrons number and the statistical fluctuation testing range. The variation of the temperature can be derived from the fact that the lighter initial isotopes of the U, Pa, Th and Ac series under the same fission conditions must be "hotter" than heavier. Really, the mass of difference of the isotopes in the same series varies up to 4%, and between the different elements within 7%.

Our data are shown in Figure 5 for several outside (depleted/enriched by neutrons) isotopes in the U, Pa, Th and Ac series, see Figure 4. The spread of the theoretical data for the same proton number is caused by the statistical fluctuation of thermodynamical parameters. We fit the theoretical data with one, two and three Gaussian functions (solid lines in Figure 5) in order to emphasize their hump structure.

One can see that suggested method qualitatively reflects all the main experimentally observed structures of the nuclear-charge spectra. The U series demonstrates a strongly expressed two-hump structure. The Pa series exhibits the transition from one- to two-hump structure of the charge spectrum. The same trend we obtain for the isotopes of the Th series. However, we have also found a weakly expressed third-hump structure in the middle of the charge spectrum of $^{229}$Th (see Figure 5c). Finally, the three-hump structure is more expressed for the 226 isotope of the Ac series.

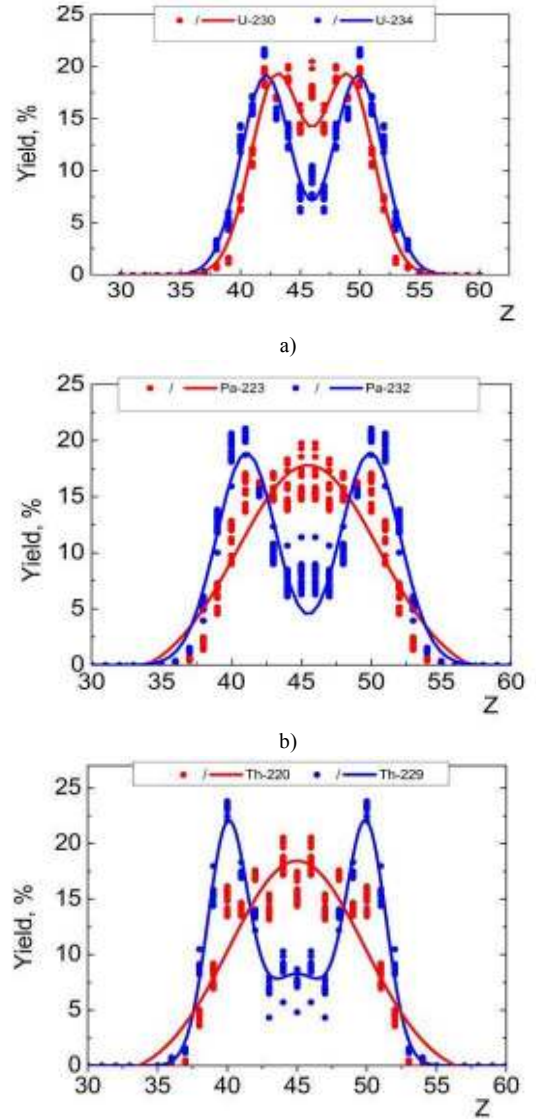

a)

b)

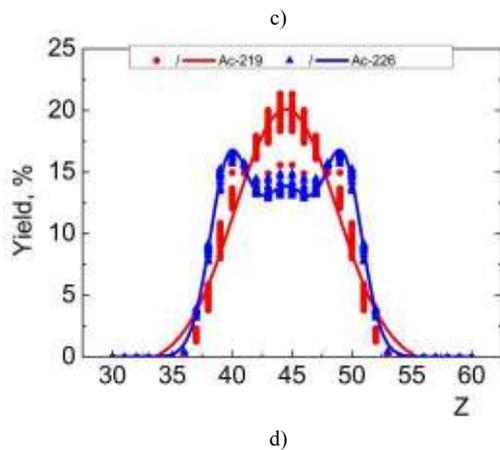
c)

d)

Figure 5. (Color online) The evaluated fission-fragment nuclear-charge spectra of the outside (depleted/enriched by neutrons) isotopes of the U → a), Pa → b), Th → c), Ac → d) series, see Figure 4

## 4. Conclusions

Thus, the statistical approach for the post-scission fragments ensembles is able to explaine the main peculiarities of the nuclear fission observed characteristics. Despite a limited number of model parameters such as isobaric constant and cumulative chain length, the proposed statistical method can explain the principal features of the mass and charge fragments spectra, their anisotropy and symmetrization with the rise of temperature.

The method allows to estimate the validity of the MCSFF data obtained with different experimental methods such as gamma-, mass spectrometry with time-of-flight mass data or the radiochemical one, e.g. with the verification of their possibilities in detecting the short-, ultra- or long-lived nuclei.

We also can assume that considering the post-fission beta$^+$ and beta$^-$ particles emission and statistical fluctuation of the binding energy will improve the analytic abilities of the proposed statistical method.

## 5. Acknowledgements

The authors are grateful to D. Symochko, V. Mazur, M. Stetc, O. Snigursky and V.V. Maslyuk for fruitful discussions and assistance. This work was carried out within the framework of the Program of informatization of the National Academy of Sciences of Ukraine and with the partial support of the Project X-5-7.